\documentclass[aps,pra,twocolumn,showpacs,superscriptaddress,eqsecnum]{revtex4}
\usepackage{epsfig}
\usepackage{graphicx}
\usepackage{bm}%bold math
\begin{document}
\date{\today}
\title{Optimal unambiguous discrimination of two subspaces as a case in mixed state discrimination}
\author{J\'{a}nos A. Bergou}
\affiliation{Department of Physics and Astronomy, Hunter College of the City University of New York, \\ 695 Park Avenue, New York, NY 10021}
\author{Edgar Feldman}
\affiliation{Department of Mathematics, Graduate Center of the City University of New York, \\ 365 Fifth Avenue, New York, NY 10016}
\author{Mark Hillery}
\affiliation{Department of Physics and Astronomy, Hunter College of the City University of New York, \\ 695 Park Avenue, New York, NY 10021}
%\begin{document}

\begin{abstract}
We show how to optimally unambiguously discriminate between two subspaces of 
a Hilbert space.
In particular we suppose that we are given a quantum system in either 
the state $|\psi_{1}\rangle$,
where $|\psi_{1}\rangle$ can be any state in the subspace $S_{1}$, 
or $|\psi_{2}\rangle$, where
$|\psi_{2}\rangle$ can be any state in the subspace $S_{2}$, and our 
task is to determine in
which of the subspaces the state of our quantum system lies.  We do not 
want to make any error,
which means that our procedure will sometimes fail if the subspaces are 
not orthogonal.  This is
a special case of the unambiguous discrimination of mixed states.  
We present the POVM that
solves this problem and several applications of this procedure, 
including the discrimination of
multipartite states without classical communication.
\end{abstract}
\maketitle

\section{Introduction}\label{section1}

The discrimination of quantum states is an area that has received 
considerable attention. For a recent review see \cite{springer}, for example. 
Pure state discrimination has proven to be useful for both quantum 
cryptography and for quantum algorithms \cite{bennett,bergou1}.  The basic 
protocol is the following.  One has a list of possible quantum
states, and one is presented with a system that is guaranteed to be in 
one of them.  Our task is
to determine which.  If the states are not orthogonal, they cannot be 
discriminated perfectly,
and it is necessary to specify what kind of strategy we wish to use.  One 
possibility is
minimum error discrimination \cite{helstrom}.  In this procedure, we always 
specify a state, but, because the
states are not orthogonal, there is some chance that we will make a mistake.
Minimum-error procedures minimize the chance of making a mistake.  Another 
alternative is
unambiguous discrimination.  This procedure never makes a mistake, but it can 
sometimes fail.
If the list of states contains $N$ possibilities, then this procedure has
$N+1$ possible outputs, one for each 
of the states on list plus a failure output.  If we receive an output that 
specifies one of the states, then we
know what the state of our system was, and if we receive the failure output, 
then we have no
idea what it was. An optimum unambiguous procedure is one that minimizes the 
probability of
receiving the failure output.  We shall be considering unambiguous 
discrimination here.

The problem of optimal unambiguous discrimination between two pure quantum states was solved by Ivanovic, Dieks and Peres \cite{ivanovic,dieks,peres}.  They assumed that each of the two states were equally probable.  The case in which they are not was treated by Jaeger and Shimony \cite{jaeger} (for a somewhat simpler derivation of their result see \cite{bergou2}).  The case of more than two pure states is not simple and has been considered by a number of authors.  There are a few general results, but explicit procedures are available only for special cases \cite{terno,sun1}.
One important result is that for unambiguous discrimination to be possible, the states on the list must be linearly independent \cite{chefles}.  There are also lower bounds on the failure probability \cite{zhang}.

The discrimination of two mixed states has only been considered more recently.
Before we cite the previous results it will be useful to introduce some 
terminology at this point. The support $S_{j}$ of the density operator 
$\rho_{j}$, describing a quantum state, is the subspace of the entire 
Hilbert space $\mathcal{H}$ spanned by the eigenvectors of $\rho_{j}$ 
belonging to nonzero eigenvalues (for  $j=1,2$). The rank of the density 
operator is equal to the dimension of its support. The subspace orthogonal 
to its support, ${\bar S}_{j}$, is the kernel of the density operator 
$\rho_{j}$ such that $\mathcal{H}=S_{j} \oplus {\bar S}_{j}$. We shall 
denote the projector onto $S_{j}$ by $P_{j}$ and the projector onto 
${\bar S}_{j}$ by ${\bar P}_{j}$. Equipped with these definitions we can 
now make the following general statement. For two mixed quantum states 
unambiguous discrimination is possible with a finite probability of 
success if and only if the supports of their density operators are not 
identical. Indeed, in such a case the kernel of at least one of them is 
not empty and a projective measurement along this kernel unambiguously 
identifies the other state. On the other hand, if the supports are 
identical, then so are their kernels, 
and there is no direction in ${\mathcal H}$ that 
could unambiguously identify at least one of the density operators. 
  
We can now return to a listing of earlier results. The POVM for unambiguously 
discriminating between a pure state and a rank two mixed state was derived in
\cite{sun2} and subsequently generalized to the case of unambiguously 
discriminating a pure
state from any mixed state \cite{bergou1}.  Lower bounds on the failure 
probability for the unambiguous discrimination of two mixed states were 
derived by Rudolph \emph{et al}.
\cite{rudolph} and for an arbitrary number of mixed states by 
Feng \emph{et al}. \cite{feng}.
Raynal \emph{et al}. \cite{raynal} proved two theorems that make it 
possible to reduce the problem of
unambiguously discriminating between two arbitrary mixed states of rank 
$k_{1}$ and $k_{2}$ to the discrimination of
two states of the same rank, $k \leq min(k_{1},k_{2})$, in a $2k$-dimensional 
space.  Building on the results of \cite{rudolph} and \cite{feng}, Herzog and Bergou \cite{HB} found  explicit solutions for some special cases along with necessary conditions for the 
saturation of the lower bound. In particular, these results showed that unlike 
in the case of two pure states whether or not the lower bound can be attained 
depends not only on the value of the prior probabilty of the states but also 
on their structure. There are mixed states for which the lower bound can not 
be reached for any value of the prior probability. In \cite{raynal2} the 
optimal measurement operators were constructed explicitly for some special 
cases.   

In this paper we shall consider the unambiguous discrimination between two 
subspaces.  What this means is the following.  A state is chosen from one
of two subspaces, and we wish to determine to which of the subspaces the
state belongs.  Within each subspace each state is equally likely, though
one subspace may be more likely than the other.  One place in which this 
type of problem has arisen is in the consideration of programmable 
discriminators \cite{bergou3}.  In this case, one is given three qubits, 
the first two are
arbitrary but the third is guaranteed to be identical to either the first
or the second qubit, and the problem is to determine which two qubits
are identical.  The problem
can be solved by realizing that one is, in fact, discriminating between
two subspaces, the first being the subspace of three-qubit states that is
symmetric in the first and third qubits, and the second being the subspace
that is symmetric in the second and third qubits.

Subspace discrimination is a special case of the discrimination of two 
mixed states; in this case the density matrices
are just proportional to the projection operators onto the subspaces.  Making use of the results of Raynal \emph{et al}. \cite{raynal}, we can restrict our attention to the case of two subspaces of dimension $k$
in a $2k$ dimensional space.  In particular, let $S_{1}$, and $S_{2}$ be $k$-dimensional subspaces of the entire Hilbert space, $\mathcal{H}$, which has dimension $2k$.
We can assume that the intersection of $S_{1}$ and $S_{2}$ is just the zero vector, a situation  which we henceforth refer to as general position.
We assume that $\rho_{1}= (1/k)P_{1}$ occurs with probability $\eta$, and
$\rho_{2}= (1/k)P_{2}$ occurs with probability $1-\eta$, where $P_{j}$ is the projection
onto $S_{j}$, for $j=1,2$.  The POVM that distinguishes them has three
elements, $\Pi_{1}$, $\Pi_{2}$, and $\Pi_{0}=I-\Pi_{1}-\Pi_{2}$, all of which are positive operators.  The probability of identifying $\rho_{j}$ if we are given $\rho_{j}$ is $p_{j}={\rm Tr}(\rho_{j}\Pi_{j})$,
and the probability of failing to identify it is $q_{j}={\rm Tr}(\rho_{j}\Pi_{0})$, for $j=1,2$.  The
condition that a state never be misidentified implies that $\Pi_{1}\rho_{2}=\Pi_{2}\rho_{1}=0$.
The average failure probability is
\begin{equation}
\label{Qave}
Q=\eta q_{1}+(1-\eta )q_{2} \ ,
\end{equation}
and our object is to find, for a given $\eta$,  a POVM that minimizes $Q$.  From the results in
\cite{rudolph} and \cite{feng}, we have that
\begin{equation}
\label{fidbound}
Q\geq 2\sqrt{\eta (1-\eta )} F(\rho_{1},\rho_{2}) \ ,
\end{equation}
where the fidelity between the two density matrices is given by
\begin{equation}
F(\rho_{1},\rho_{2})=\rm{Tr}((\rho_{1}^{1/2}\rho_{2}\rho_{1}^{1/2})^{1/2}) \ .
\label{fidelity}
\end{equation}
The conditions, under which this bound can be saturated, have been investigated in \cite{HB}. In \cite{raynal2} the optimum POVM has been given explicitly for certain special cases. 
As we shall see, unlike for the case of two pure states, this bound cannot always be reached for two mixed states, in agreement with \cite{HB}.  The optimal measurement procedure
depends on the value of $\eta$,  a feature it has in common with the procedure for discriminating two pure states. For two pure states there is always a range of $\eta$ where equality in (\ref{fidbound}) can be reached. In the case of two mixed states, however, the optimal measurement procedure also depends on the structure of the two density operators, a distinctive  feature that has no equivalent in the case of two pure states. For $\eta$ near $0$ or $1$ the optimal measurements are projective ones.
In the intermediate regime, the optimal measurements are intermittently POVM's or projective measurements and, in general, their failure probability is higher than the fidelity bound (\ref{fidbound}). Only under very special conditions shall we find that the fidelity bound can be saturated.

The main technical device that we shall use to find the optimal measurements is that of Jordan
bases.  These bases take the following form.  The states $|\psi_{1}\rangle , \ldots |\psi_{k}\rangle$
form an orthonormal basis for $S_{1}$, $|\psi_{k+1}\rangle ,\ldots |\psi_{2k}\rangle$ form an orthonormal basis for $S_{2}$, and, in addition these states have the property that
\begin{equation}
\langle\psi_{i}|\psi_{k+j}\rangle = \delta_{ij}\cos\theta_{i} ,
\end{equation}
where $\cos\theta_{1}\geq\cos\theta_{2}\geq\ldots\cos\theta_{k}$, and $1\leq i,j\leq k$.  The
states $|\psi_{1}\rangle , \ldots, |\psi_{k}\rangle$ and $|\psi_{k+1}\rangle ,\ldots, |\psi_{2k}\rangle$
are called Jordan bases and the angles $\theta_{i}$ are called the Jordan angles.  Bases
satisfying these conditions can be constructed for any two subspaces \cite{jordan}.   Note that
the basis vectors $|\psi_{i}\rangle$ and $|\psi_{k+i}\rangle$ are eigenvectors of the operators
$P_{1}P_{2}P_{1}$ and $P_{2}P_{1}P_{2}$, respectively, where
\begin{eqnarray}
P_{1}P_{2}P_{1} |\psi_{i}\rangle & = & \cos^{2}\theta_{i}|\psi_{i}\rangle \ ,  \nonumber \\
P_{2}P_{1}P_{2}|\psi_{k+i}\rangle & = & \cos^{2}\theta_{i}|\psi_{k+i}\rangle \ ,
\end{eqnarray}
where $1\leq i \leq k$.

The paper is organized as follows. Sec. \ref{section2} is devoted to the 
derivation of the general results. Besides giving the general theory for 
distinguishing two $k$ dimensional subspaces in a $2k$ dimensional Hilbert 
space, these results also hold for a rather general class of density 
operators, so they are directly relevant to the problem of optimal 
unambiguous discrimination between two mixed states. In Sec. \ref{section3} 
we present some possible applications of the results. 
Finally, in Sec. \ref{section4} we give a brief summary and 
outlook for future research. 

\section{Distinguishing subspaces and its relation to the discrimination of mixed states}\label{section2}

We shall actually solve a somewhat more general problem than the 
discrimination of two subspaces.  Let $S_{1}$ and $S_{2}$ be $k$ dimensional 
subspaces of a $2k$
dimensional complex Hilbert space $\mathcal{H}$ which are in general 
position, as discussed in the introduction, and
let $\{|\psi_{1}\rangle,\ldots,|\psi_{k}\rangle \}$ and 
$\{|\psi_{k+1}\rangle,\dots,|\psi_{2k}\rangle\}$ be Jordan bases associated 
to these subspaces.  Consider the two density matrices
\begin{eqnarray}
\rho_{1}&=&\sum_{i=1}^{k}\alpha_{i}|\psi_{i}\rangle\langle\psi_{i}| \ , 
\nonumber \\
\rho_{2}&=&\sum_{i=1}^{k}\beta_{i}|\psi_{i+k}\rangle\langle\psi_{i+k}| \ ,
\end{eqnarray}
where $\alpha_{i}>0$, $\sum_{i}\alpha_{i}=1$, $\beta_{i}>0 $, and 
$\sum_{i}\beta_{i}=1$.
Clearly, $\rho_{1}$ has support in $S_{1}$ and $\rho_{2}$ has support 
in $S_{2}$.  In this
case the orthonormal frames for the density matrices given by the spectral 
theorem coincide with the Jordan frames of the supports. If
\begin{eqnarray}
\rho_{1} &  = & (1/k)\sum_{i=1}^{k}|\psi_{i}\rangle\langle\psi_{i}| 
\nonumber \\
\rho_{2} &  = & (1/k)\sum_{i=1}^{k}|\psi_{i+k}\rangle\langle\psi_{i+k}| ,
\end{eqnarray}
we say that  $\rho_{1}$ and $\rho_{2}$ are uniformly mixed states, and 
discriminating between uniformly mixed states corresponds to the case of 
discriminating
between the subspaces $S_{1}$ and $S_{2}$.  Inserting the weights 
$\alpha_{i}$ and
$\beta_{i}$ will allow us to address several issues in the general 
theory of mixed state discrimination.

We will now construct an optimal POVM to unambiguously discriminate between $\rho_{1}$ and $\rho_{2}$.  The POVM elements are $\Pi_{1}$, $ \Pi_{2}$ and $\Pi_{0}=I-\Pi_{1}-\Pi_{2}$ with the properties as discussed in the introduction.  $\Pi_{1}$ and $\Pi_{2}$ are self adjoint operators supported on ${\bar S}_{2}$ and
${\bar S}_{1}$, respectively.  In order for them to form an optimal POVM they must be positive and, crucially, the rank of $\Pi_{0}$ must not exceed $k$ \cite{raynal}.  We wish to choose $\Pi_{1}$ and $\Pi_{2}$ so that the expression
$P(\eta )=\eta {\rm Tr}(\Pi_{1} \rho_{1})+(1 - \eta){\rm Tr}(\Pi_{2}\rho_{2})$ is maximized. This is the same as minimizing the average failure probability $Q(\eta)=\eta {\rm Tr}(\Pi_{0}\rho_{1})+(1-\eta ){\rm Tr}(\Pi_{0}\rho_{2})$ that was introduced in (\ref{Qave}).

Let $T_{i}$ be the linear subspace
spanned by $|\psi_{i}\rangle$ and $|\psi_{i+k}\rangle$. The
$T_{i}$'s, with $1\leq i\leq k $, are mutually orthogonal, two dimensional
subspaces, which are invariant under both $\rho_{1}$ and $\rho_{2}$.
Let us define states $|z_{i}\rangle$ and $|y_{i}\rangle$ as
\begin{eqnarray}
|\psi_{i}\rangle & = & \sin\theta_{i}|z_{i}\rangle +
\cos\theta_{i}|\psi_{i+k}\rangle \nonumber \\
|\psi_{i+k}\rangle & = & \sin\theta_{i}|y_{i}\rangle +
\cos\theta_{i}|\psi_{i}\rangle  , 
\end{eqnarray}
where $|z_{i}\rangle$ $(|y_{i}\rangle)$
is the orthogonal complement of $|\psi_{i+k}\rangle$
$(|\psi_{i}\rangle)$ in $T_{i}$. Furthermore
$ \{ |z_{1}\rangle,\ldots ,|z_{k}\rangle \}$ forms an orthonormal basis for
${\bar S}_{2}$, and
$\{ |y_{i}\rangle,\ldots ,|y_{k}\rangle \}$ forms an orthonormal basis for
${\bar S}_{1}$.

If we write $\Pi_{1}=\sum_{i,j=1}^{k} a_{ij}|z_{i}\rangle\langle z_{j}|$ and
$\Pi_{2}=\sum_{i.j=1}^{k} b_{ij}|y_{i}\rangle\langle y_{j}|$,  then
\begin{equation}
p_{1}={\rm Tr}(\Pi_{1}\rho_{1})=\sum_{i=1}^{k} a_{ii} \alpha_{i} 
\sin^{2} \theta_{i} \ ,
\end{equation}
and
\begin{equation}
p_{2}={\rm Tr}(\Pi_{2}\rho_{2})=\sum_{i=1}^{k} b_{ii} \beta_{i} 
\sin^{2} \theta_{i} \ .
\end{equation}
These equations do not depend upon the off-diagonal terms of $\Pi_{1}$ and $\Pi_{2}$.
If $\Pi_{1}$ and $\Pi_{2}$ are to be elements of a POVM they must be positive
so $a_{ii}, b_{ii} \geq 0$ is a minimum requirement. The presence of off-diagonal elements imposes additional restrictions on the diagonal elements if we wish to ensure positivity. Since the off-diagonal
 elements do not play a role in $p_{1}$ and $p_{2}$, it suffices to search for our optimal POVM among the
  diagonal operators.

Let ${\bar p}_{i}=\langle\psi_{i}|\Pi_{1}|\psi_{i}\rangle$ and
 ${\bar p}_{i+k}=\langle \psi_{i+k}|\Pi_{2}|\psi_{i+k}\rangle$ be the individual success probabilities of
 the Jordan basis states for $1\leq i \leq k$. Let
${\bar q}_{i}=\langle\psi_{i}|\Pi_{0}|\psi_{i}\rangle$ and
 ${\bar q}_{i+k}=\langle\psi_{i+k}|\Pi_{0}|\psi_{i+k}\rangle$ be the individual failure probabilities of
 the Jordan basis states for $1 \leq i \leq k$. Here we introduced the overbar notation in order to
 distinguish the partial success and failure probabilities, ${\bar p}_{i}$ and ${\bar q}_{i}$, of
  $|\psi_{i}\rangle$ from the total success and failure probabilities, $p_{j}$ and $q_{j}$, of $\rho_{j}$
  for $i \leq k$ and $j=1,2$. Obviously, ${\bar p}_{i} + {\bar q}_{i} = 1$ 
holds.  We can then express the POVM operators as $\Pi_{1}=\sum_{i=1}^{k}
\Pi_{1,i}$ and $\Pi_{2}=\sum_{i=1}^{k}\Pi_{2,i}$
where 
\begin{equation}
\Pi_{1,i}=\frac{1-\bar
q_{i}}{\sin^{2}\theta_{i}}|z_{i}\rangle\langle z_{i}| ,
\end{equation}
and
\begin{equation}
\Pi_{2,i}=\frac{1-\bar
q_{i+k}}{\sin^{2}\theta_{i}}|y_{i}\rangle\langle y_{i}| . 
\end{equation}
We can also set $\Pi_{0}=\sum_{i=1}^{k}\Pi_{0,i}$, where
$\Pi_{0,i}=I_{T_{i}}- \Pi_{1,i}- \Pi_{2,i}$, and $I_{T_{i}}$ is the 
identity in $T_{i}$.  

We now need to determine the values of ${\bar q}_{i}$ and ${\bar q}_{i+k}$.
This can be done by noticing that what we have done is to reduce our problem
to $k$ problems of optimally discriminating two vectors,
in particular the vector $|\psi_{i}\rangle$ from the vector
$|\psi_{i+k}\rangle$ in $T_{i}$.  In more detail the situation is the 
following.  In our overall ensemble, with $\rho_{1}$ occurring with
probability $\eta$ and $\rho_{2}$ with probability $1-\eta$, 
the probability of occurrence for $|\psi_{i}\rangle$ is 
$\eta\alpha_{i}$ and the probability for the occurrence of $|\psi_{i+k}\rangle$
is $(1-\eta )\beta_{i}$.  Therefore, the probability for the occurrence 
of a vector in $T_{i}$ is just the sum of these probabilities,
\begin{equation}
p(T_{i})=\eta\alpha_{i}+(1-\eta )\beta_{i} 
\end{equation}
Now, the probability that $|\psi_{i}\rangle$ occurs given that $T_{i}$ has 
ocurred is $p(i|T_{i})=\eta\alpha_{i}/p(T_{i})$ and the probability
that $|\psi_{i+k}\rangle$ occurs given that $T_{i}$ has occurred is
$p(i+k|T_{i})=(1-\eta )\beta_{i}/p(T_{i})$. Consequently, in $T_{i}$ 
we want to unambiguously discriminate
$|\psi_{i}\rangle$ occurring with probability $p(i|T_{i})$ and 
$|\psi_{i+k}\rangle$ occuring with probability $p(i+k|T_{i})$ so as to 
minimize the failure probability
\begin{equation}
\label{Qi}
Q_{i}(\eta )=p(i|T_{i}){\bar q}_{i}+p(i+k|T_{i}){\bar q}_{i+k} .
\end{equation}
This problem was first solved by Jaeger and Shimony \cite{jaeger}, (a
somewhat simpler solution is given in \cite{bergou2}), and we can now make use
of that solution.  Let
\begin{equation}
I_{i}= \left[ \frac{\beta_{i} \cos ^{2} \theta_{i}}{\alpha_{i}+
\beta_{i} \cos^{2}\theta_{i}}, \frac{\beta_{i}}{\beta_{i} +
\alpha_{i} \cos^{2}\theta_{i}} \right]  = [c_{i},d_{i}]
\end{equation} 
and, in addition, let
\begin{equation}
 {\bar q}_{i}^{opt} (\eta )=\sqrt{\frac{(1-\eta )\beta_{i}}{\eta \alpha_{i}}}
\cos \theta_{i} \ .
\end{equation}
For a given $\eta$, the value of $\bar q_{i}$ which
minimizes $Q_{i}(\eta)$ is
\begin{equation}
\label{qiopt}
q_{i}( \eta )=\left\{\begin{array}{ll} 1 & {\eta \leq c_{i}}\\
 {\bar q}_{i}^{opt} (\eta) & \mbox{ if $ \eta $ is in $I_{i}$}\\
 \cos ^{2}\theta_{i}& {\eta\geq d_{i}}\end{array} \right.
\end{equation}
and $q_{k+i}(\eta )=\cos^{2}\theta_{i}/q_{i}(\eta )$. Furthermore the rank of
$\Pi_{0,i}$ is one, with the nonzero eigenvalue given by
\begin{equation} \lambda_{i}= \left(
\frac{\cos^{2}\theta_{i}}{{\bar q}_{i}} -2 \cos^{2} \theta_{i} +
{\bar q}_{i} \right) \frac{1}{\sin^{2}\theta_{i}} \ ,
\end{equation} 
with the corresponding eigenstate
\begin{equation}
\label{zeta1} |\zeta_{i}\rangle=\frac{\cos\theta_{i}(1-{\bar
q}_{i})}{\sin^2\theta_{i}} |\psi_{i+k}\rangle +
  \frac{{\bar q}_{i}-\cos^{2}\theta_{i}}{\sin^{2}\theta_{i}} 
|\psi_{i}\rangle \ .
\end{equation} 
This specifies the POVM within $T_{i}$.

The optimal overall failure probability can now be expressed as
\begin{eqnarray}
Q^{opt} & = & \sum_{i=1}^{k} Q_{i}^{opt}p(T_{i})  \nonumber \\
 & = & \sum_{i=1}^{k}[\eta \alpha_{i}{\bar q}_{i}(\eta) 
+(1-\eta) \beta_{i}{\bar q}_{k+i}(\eta)] \ ,
\end{eqnarray}
where $Q_{i}^{opt}$ is the failure probability that results when 
Eq.\ (\ref{qiopt}) is substituted into Eq.\ (\ref{Qi}).
Its explicit expression is given by
\begin{equation}
Q_{i}^{opt}p(T_{i}) = \left\{\begin{array}{ll} \eta \alpha_{i}+(1-\eta)
\beta_{i}\cos^{2}\theta_{i} & \mbox{if $\eta \leq c_{i}$}\\
 2\sqrt{\eta(1-\eta)\alpha_{i}\beta_{i}}|\cos\theta_{i}| & \mbox{if $c_{i} \leq \eta \leq d_{i}$}\\
 \eta\alpha_{i}\cos ^{2}\theta_{i}+(1-\eta)\beta_{i} & \mbox{if $\eta \geq d_{i}$}\end{array} \right. \ .
\label{Qiopt}
\end{equation}
The center line is the geometric mean of the two terms in either the first or the last line and, therefore, represents an absolute minimum for $Q_{i}$.  We obtain the absolute possible minimum of the total failure probability if we sum the center lines for all $1 \leq i \leq k$. The summation yields
\begin{equation}
\label{Qopt}
Q^{opt} = 2\sqrt{\eta(1-\eta)} \sum_{i=1}^{k}
\sqrt{\alpha_{i}\beta_{i}}|\cos\theta_{i}| \ .
\end{equation}
Clearly, this absolute minimum can only be realized if and only if the intersection of all of the intervals $I_{i}$ is not empty and the operating value of $\eta$ is in this intersecton. 

The interpretation of this expression is straightforward. Making use of 
the structure of the Jordan bases, we obtain $\rho_{1}^{\frac{1}{2}}
\rho_{2}\rho_{1}^{\frac{1}{2}} = \sum_{i}\alpha_{i}\beta_{i} 
\cos^{2}\theta_{i}|\psi_{i}\rangle\langle \psi_{i}|$, after a simple 
calculation. Comparing this expression with Eq.\ (\ref{fidelity}) tells us  
immediately that Eq.\ (\ref{Qopt}) can be cast to the form
\begin{equation}
 Q^{opt}= 2\sqrt{\eta (1-\eta)}F(\rho_{1},\rho_{2}) \ ,
\end{equation}
where $F(\rho_{1},\rho_{2}) = \sum_{i}\sqrt{\alpha_{i}\beta_{i}}
|\cos\theta_{i}|$. Here $F(\rho_{1},\rho_{2})$ is the fidelity between 
$\rho_{1}$ and $\rho_{2}$, constructively proving that the fidelity bound, 
Eq.\ (\ref{fidbound}), can be saturated. To obtain an explicit expression 
for the fidelity would be a hopeless task, in general. What made it possible 
here is the fact that the density operators are diagonal in the Jordan bases 
of their support and we could take full advantage of the ensuing Jordan 
structure. Furthermore, the above expression for the optimum failure 
probability holds only if $\eta$ is an element of the intersection of all 
of the $I_{i}$ intervals, $I_{0}=\bigcap_{i=1}^{k}I_{i}$. $I_{0}$ may be 
empty, it often is. Note, however, that in the case where $\rho_{1}$ and 
$\rho_{2}$ are uniformly mixed states, which is the case of subspace 
discrimination,
\begin{equation}
I_{i}=\left[ \frac{\cos^{2}\theta_{i}}{1+\cos^{2}\theta_{i}},
\frac{1}{1+\cos^{2}\theta_{i}} \right]
\subseteq I_{i+1}
\end{equation}
so $I_{0}=I_{1} \neq \emptyset$ and the fidelity result holds in the 
entire $I_{1}$ interval.

For the case when $k = 2$ there are only two such intervals, $I_{1}$ and $I_{2}$, and  we will now give a complete classification of their intersection pattern. To this end, we first introduce a one-parameter characterization of $\rho_{1}$ and $\rho_{2}$. Let us set $\alpha_{1}=\alpha$ and, consequently, $\alpha_{2}=1-\alpha$.  Similarly, we set $\beta_{1}=\beta$ and, consequently, $\beta_{2}=1-\beta$. We can now introduce a two-dimensional parameter plane, $\alpha \beta$, where the square in the first quadrant, bounded by $0 \leq \alpha \leq 1$ and $0 \leq \beta \leq 1$,  corresponds to physically acceptable choices for mixed states. So our task is reduced to finding the regions within this square with qualitatively different overlap patterns. When $\cos^{2}\theta_{1}>\cos^{2}\theta_{2}$, which is the convention that we adopted at the beginning, the patterns can be sorted into five categories: \newline
i) $d_{1}<c_{2}$, i.\ e. $I_{1}$ is to the left of $I_{2}$ and their intersection is empty. This happens when
\begin{equation}
\beta \leq {\bar \beta}_{1}(\alpha)=\frac{\alpha \cos^{2}\theta_{1}
\cos^{2}\theta_{2}}{1-\alpha(1-\cos^{2}\theta_{1}\cos^{2}\theta_{2})} \ .
\end{equation}
${\bar \beta}_{1}(\alpha)$ is a hyperbola in the $\alpha \beta$ plane 
and it is the divider between this region and the next, when \newline
ii) $c_{1}<c_{2}<d_{1}<d_{2}$, i.\ e. the right end of $I_{1}$ partially
overlaps with the  left end of $I_{2}$ and their intersection is the overlap. 
This happens when
\begin{equation}
\beta \leq {\bar \beta}_{2}(\alpha)=\frac{\alpha
\cos^{2}\theta_{2}}{\cos^{2}\theta_{1}-\alpha(\cos^{2}\theta_{1}-\cos^{2}\theta_{2})} \ .
\end{equation}
${\bar \beta}_{2}(\alpha)$ is a hyberbola in the $\alpha \beta$ plane 
and it is the divider between this region and the next, when \newline
iii) $c_{2}<c_{1}$ and $d_{1}<d_{2}$, i.\ e. $I_{1}$ is inside $I_{2}$ and the intersection coincides with $I_{1}$. This happens when
\begin{equation}
\beta \leq {\bar \beta}_{3}(\alpha)=\frac{\alpha
\cos^{2}\theta_{1}}{\cos^{2}\theta_{2}+ \alpha(\cos^{2}\theta_{1}-
\cos^{2}\theta_{2})} \ .
\end{equation}
${\bar \beta}_{3}(\alpha)$ is a hyberbola in the $\alpha \beta$ plane 
and it is the divider between this region and the next, when \newline
iv) $c_{1}<c_{2}<d_{2}<d_{1}$, i.\ e. the left end of $I_{1}$ partially overlaps with the right end of $I_{2}$ and the intersection is the overlap. This happens when
\begin{equation}
\beta \leq {\bar \beta}_{4}(\alpha)=\frac{\alpha}{\cos^{2}\theta_{1}\cos^{2}
\theta_{2} + \alpha(1-\cos^{2}\theta_{1}\cos^{2}\theta_{2})} \ .
\end{equation}
${\bar \beta}_{4}(\alpha)$ is a hyberbola in the $\alpha \beta$ plane and it 
is the divider between this region and the next, when, finally \newline
v) $d_{2}<c_{1}$, i.\ e. $I_{1}$ is to the right of $I_{2}$ and the intersection is empty. This happens when
\begin{equation}
\beta \geq {\bar \beta}_{4} \ .
\end{equation}

\begin{figure}[!ht]
\epsfig{file=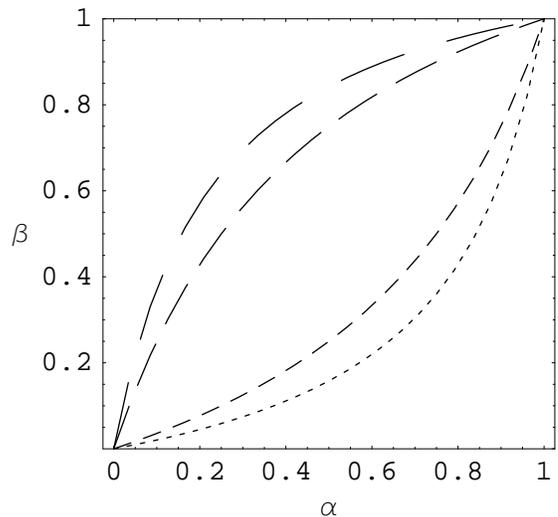, height=7cm}
\caption{Regions of the parameter space. Dotted line: 
${\bar \beta}_{1}(\alpha)$, short dashed line: ${\bar \beta}_{2}(\alpha)$, 
medium dashed line: ${\bar \beta}_{3}(\alpha)$, long dashed line: 
${\bar \beta}_{4}(\alpha)$. In the region below ${\bar \beta}_{1}(\alpha)$ 
and above ${\bar \beta}_{4}(\alpha)$ the intersection of $I_{1}$ and $I_{2}$ 
is empty, the fidelity bound can not be reached; in the regions between 
${\bar \beta}_{1}(\alpha)$ and ${\bar \beta}_{2}(\alpha)$, and between 
${\bar \beta}_{3}(\alpha)$ and ${\bar \beta}_{4}(\alpha)$ the intervals 
$I_{1}$ and $I_{2}$ partially overlap, the fidelity bound can be reached 
in these overlaps; and in the region between ${\bar \beta}_{2}(\alpha)$ 
and ${\bar \beta}_{3}(\alpha)$ the interval $I_{1}$ is inside $I_{2}$, 
the fidelity bound can be reached in the entire $I_{1}$. For the figure we 
used the values $\cos \theta_{1} = \frac{\sqrt{3}}{2}$ and 
$\cos \theta_{2} = \frac{1}{2}$.}
\label{Fig1}
\end{figure}

We note that when $\cos^{2}\theta_{1} = \cos^{2}\theta_{2}$ the two 
inner dividers, ${\bar \beta}_{2}$ and ${\bar \beta}_{3}$, both 
degenerate into the diagonal of the square, $\beta = \alpha$. Our 
findings are summarized in Fig. \ref{Fig1} where the five regions of 
the parameter space, resulting from the four dividers ${\bar \beta}_{1},
\ldots,{\bar \beta}_{4}$, are displayed for the representative 
values $\cos^{2}\theta_{1} = \frac{3}{4}$ and $\cos^{2}\theta_{2} 
= \frac{1}{4}$.

Next we give two illustrative examples for $k=2$. In these examples we prescribe $\Pi_{0}$ by giving its eigenvectors $|\zeta_{i}\rangle$ and eigenvalues $\lambda_{i}$ for the possible $\eta$ interval configurations. We adopt the notation $|\zeta_{i}(\eta)\rangle$ ($\lambda_{i}(\eta)$) for the eigenvector (eigenvalue) corresponding to  ${\bar q}_{i}(\eta) = {\bar q}^{opt}_{i}$.

In our first example we consider the case where $\rho_{1}$ and $\rho_{2}$ are uniformly mixed, corresponding to subspace discrimination. Then $I_{1}=[c_{1},d_{1}] \subseteq I_{2}=[c_{2},d_{2}]$, where
\begin{equation}
c_{i}=\frac{\cos^{2}\theta_{i}}{1+\cos^{2}\theta_{i}} \ , \hspace{0.5cm} d_{i}=\frac{1}{1+\cos^{2}\theta_{i}} \ .
\end{equation}
The interval $0 \leq \eta \leq 1$ is divided into five subintervals, $[0,c_{2}]$, $[c_{2},c_{1}]$, $[c_{1},d_{1}]$, $[d_{1},d_{2}]$, and $[d_{2},1]$. Table \ref{table1} summarizes the behavior of $\Pi_{0}$ in each of the subintervals.
\begin{table}[h!]
\caption{Eigenvalues and eigenvectors of $\Pi_{0}$ in the various intervals of $\eta$ for Example 1. Both states are uniformly mixed and the intersection of $I_{1}$ and $I_{2}$ is the entire $I_{1}$. The fidelity bound can be reached in the intersection.}
\begin{tabular}{||c||c|c|c|c||}
\hline
$\eta$ & $\lambda_{1}$ & $\lambda_{2}$ &  $|\zeta_{1}\rangle$ & $|\zeta_{2}\rangle$ \\ \hline \hline
$[0,c_{2}]$ & $1$  & $1$  & $|\psi_{1}\rangle$ & $|\psi_{2}\rangle$  \\ \hline
$[c_{2},c_{1}]$& $1$ & $\lambda_{2}(\eta)$  & $|\psi_{1}\rangle$ & $|\zeta_{2}(\eta)\rangle$ \\ \hline
$[c_{1},d_{1}]$ & $\lambda_{1}(\eta)$ & $\lambda_{2}(\eta)$  & $|\zeta_{1}(\eta)\rangle$ & $|\zeta_{2}(\eta)\rangle$ \\ \hline
$[d_{1},d_{2}]$ & $1$ & $\lambda_{2}(\eta)$ & $|\psi_{3}\rangle$ & $|\zeta_{2}(\eta)\rangle$ \\ \hline
$[d_{2},1]$ & $1$ & $1$ & $|\psi_{3}\rangle$ & $|\psi_{4}\rangle$ \\ \hline
\end{tabular}
\\
\label{table1}
\end{table}

For our second example we choose $ \rho_{1}$ and $\rho_{2}$ so that 
$d_{1}\leq c_{2}$, which can be easily arranged by picking $\beta < 
{\bar \beta}_{1}(\alpha)$. We then have  intervals $[0,c_{1}]$, 
$[c_{1},d_{1}]$, $[d_{1},c_{2}]$, $[c_{2},d_{2}]$, and $[d_{2},1]$.  This 
is a situation in which the fidelity bound for the failure probability, 
Eq.\ (\ref{fidbound}), can never be achieved. The behavior of $\Pi_{0}$  
is summarized in Table \ref{table2}.
\begin{table}[h!]
\caption{Eigenvalues and eigenvectors of $\Pi_{0}$ in the various intervals of $\eta$ for the second example in the text. The intersection of $I_{1}$ and $I_{2}$ is empty and the fidelity bound cannot be reached in this case.}
\begin{tabular}{||c||c|c|c|c||}
\hline
$\eta$ & $\lambda_{1}$ & $\lambda_{2}$ &  $|\zeta_{1}\rangle$ & $|\zeta_{2}\rangle$ \\ \hline \hline
$[0,c_{1}]$ & $1$  & $1$  & $|\psi_{1}\rangle$ & $|\psi_{2}\rangle$  \\ \hline
$[c_{1},d_{1}]$& $\lambda_{1}(\eta)$ & $1$  & $|\zeta_{1}(\eta)\rangle$ & $|\psi_{2}\rangle$ \\ \hline
$[d_{1},c_{2}]$ & $1$ & $1$ & $|\psi_{3}\rangle$ & $|\psi_{2}\rangle$ \\ \hline
$[c_{2},d_{2}]$ & $1$ & $\lambda_{2}(\eta)$ & $|\psi_{3}\rangle$ & $|\zeta_{2}(\eta)\rangle$ \\ \hline
$[d_{2},1]$ & $1$ & $1$ & $|\psi_{3}\rangle$ & $|\psi_{4}\rangle$ \\ \hline
\end{tabular}
\\
\label{table2}
\end{table}

The trend is clear from these two examples. For each of the five regions of the $\alpha\beta$ parameter plane the operating value of $\eta$ can have five possibilities. It can be outside of the intervals $I_{1}$ and $I_{2}$ (three such intervals if $I_{1}$ and $I_{2}$ do not intersect and two if they do), it can be in the nonoverlapping regions of $I_{1}$ and $I_{2}$ (two such intervals) and, finally, it can be in the intersection of $I_{1}$ and $I_{2}$ (zero or one such interval). The five parameter regions and the five possibilities for $\eta$ in each of these regions give us altogether twenty five characteristically different cases. In only three of them can the fidelity bound be reached. In twelve cases the optimum measurement is a standard von Neumann projection and in the remaining ten cases it is a combination of projections in some dimensions and POVMs in the others.

We believe that these trends are general and they hold for the 
discrimination of any Rank $2$ mixed states not just the ones where 
the Jordan basis coincides with the spectral representation but it will 
be much harder to find explicitly the different regions in the parameter space and 
the different intervals of the prior probability $\eta$. We also conjecture 
that for the discrimination of two Rank $N$ mixed states there are 
$(2N+1)^{N}$ cases altogether and the fidelity bound can be reached in 
only $2N-1$ of them. Since the growth in the number of possibilities 
as a function of $N$ is faster than exponential it seems as though it 
would be extremely difficult to give a complete classification of 
the cases for $N > 2$. Furthermore, since the number of cases when the 
fidelity bound can be reached grows only linearly with $N$, the weight of 
the density operators for which the fidelity bound can be attained quickly 
becomes negligible with increasing $N$. 

\section{Applications}\label{section3}
Let us now consider a simple example that we will be able to use as 
the basis for applications
of subspace discrimination.  Let $\mathcal{H}$ be a four-dimensional 
space with the orthonormal
basis $\{ |j\rangle\ |\ j=0,\ldots, 3\}$.  For the first subspace, 
$S_{1}$, we choose the span of the
vectors $|0\rangle$ and $|1\rangle$, and for the second, $S_{2}$ 
we choose the span of the
vectors $|u_{0}\rangle = (|0\rangle +|2\rangle )/\sqrt{2}$ and 
$|u_{1}\rangle = (|1\rangle +
|3\rangle )/\sqrt{2}$.  The states $\{ |0\rangle ,|1\rangle \}$ 
and $\{ |u_{0}\rangle , |u_{1}\rangle \}$
form Jordan bases for the subspaces $S_{1}$ and $S_{2}$.  Defining the vectors
\begin{eqnarray}
|y_{1}\rangle = |2\rangle &\hspace{3mm}& |y_{2}\rangle  = |3\rangle 
\nonumber \\
|z_{1}\rangle = \frac{1}{\sqrt{2}}(|0\rangle - |2\rangle ) 
&\hspace{3mm}& |z_{2}\rangle =
\frac{1}{\sqrt{2}}(|1\rangle - |3\rangle ) ,
\end{eqnarray}
we have that ${\bar S}_{1}$ is the span of $|y_{1}\rangle$ and $|y_{2}\rangle$, and
${\bar S}_{2}$ is the span of $|z_{1}\rangle$ and $|z_{2}\rangle$.  Application of the formulas
in the previous section gives us the POVM for discriminating between $S_{1}$ and $S_{2}$.
The POVM exists for $1/3\leq \eta \leq 2/3$, where the detection operators are given by
\begin{eqnarray}
\Pi_{1} & = & \sqrt{2}\left( \sqrt{2}-\sqrt{\frac{1-\eta}{\eta}}\right) {\bar P}_{2} \nonumber \\
\Pi_{2} & = & \sqrt{2}\left( \sqrt{2} - \sqrt{\frac{\eta}{1-\eta}}\right) {\bar P}_{1}  \ .
\end{eqnarray}
The corresponding failure probability is $Q=\sqrt{2\eta (1-\eta )}$. We also mention here that this solution was already derived in \cite{HB} using a slightly less general approach.

One possible application of this POVM is the following.  Suppose that Alice and Bob cannot
communicate with each other, but they can communicate with Charlie.  Charlie wants Alice
and Bob to share a secure bit string.  He sends to Alice and Bob one particle each from either of
the two-particle states
\begin{eqnarray}
|\Psi_{0}\rangle & = & \frac{1}{\sqrt{2}}(|0\rangle_{a}|1\rangle_{b} + |1\rangle_{a}|0\rangle_{b})
\nonumber \\
|\Psi_{1}\rangle & = & \frac{1}{\sqrt{2}}(|u_{0}\rangle_{a}|u_{1}\rangle_{b} + |u_{1}\rangle_{a}
|u_{0}\rangle_{b}) .
\end{eqnarray}
If Alice and Bob both succeed in identifying which state was sent, they share a bit,
$|\Psi_{0}\rangle$ corresponding to $0$ and $|\Psi_{1}\rangle$ corresponding to $1$.  The
reduced density matrices that Alice and Bob must distinguish are $\rho_{0}=(1/2)P_{1}$,
which results if $|\Psi_{0}\rangle$ is sent, and $\rho_{0}=(1/2)P_{2}$, which results if
$|\Psi_{1}\rangle$ is sent.  The above POVM does this optimally (we shall assume that
$\eta =1/2$).  The procedure would be the following.  Charlie sends one of the two states to
Alice and Bob (one particle to each).  They independently perform their measurements.  They
then tell Charlie whether they succeeded, and he tells each of them whether the bit is valid or
not.  The bit is valid when both Alice's and Bob's measurements succeeded, and invalid
otherwise.

The security comes from the fact that $|\Psi_{0}\rangle$ and $|\Psi_{1}\rangle$ are not orthogonal.
An eavesdropper, Eve, cannot perfectly distinguish these two states.  Her measurement procedure
will either sometimes produce errors or sometimes fail.  She must, however, send particles on to
Alice and Bob.  There is no state she can send them that will guarantee that one or both of their
measurements fail, so that sometimes Alice's and Bob's measurements will tell them that they
have received a state different from the one that Charlie sent.  By comparing some of their bits
with those of Charlie, they can tell whether this has occurred.  One possibility is that they can use
some of the invalid bits, in particular the ones for which one of the measurement succeeded and
the other did not.  For example, if Alice's measurement succeeded, then she can tell Charlie
the result of her measurement, and Charlie can see whether it corresponds to the state that he
sent.  If it does not, then they know that an eavesdropper was present.

A second example concerns operator discrimination \cite{acin}-\cite{chefles2}.
Alice starts with the two-qubit state, $|\Psi_{in}\rangle = |0\rangle
|0\rangle$.  She sends the state through one of two black boxes, each black
box performing an operation on the input state.  The first black box
performs an unknown, arbitrary single qubit rotation on the second qubit.
The second first performs an unknown, arbitrary single qubit rotation on the
second quibit and a Hadamard operation on the first, and this is followed by
sending both quibits through a C-NOT gate, with the first qubit as the 
control and the second as the target.  Alice then sends the resulting
output state to Bob, who must decide which black box Alice used.  Note that
what is being done here is the discrimination between two sets of operators;
the first black box performs an abitrary operator from the first sent and
the second black box performs an arbitrary operator from the second set. 

If the input state is sent through the first black box, the output state 
that is sent to Bob is
\begin{equation}
|\Psi_{1out}\rangle = \alpha |0\rangle |0\rangle + \beta |0\rangle |1\rangle ,
\end{equation}
where $\alpha$ and $\beta$ are unknown.  If the input state was sent through
the second black box, Bob receives the state
\begin{eqnarray}
|\Psi_{2out}\rangle  & = & \frac{1}{\sqrt{2}}[\alpha (|0\rangle |0\rangle 
+ |1\rangle |1\rangle ) \nonumber \\
 &{}& + \beta (|0\rangle |1\rangle + |1\rangle |0\rangle )  .
\end{eqnarray}
The state $|\Psi_{1out}\rangle$ lies in the subspace spanned by the vectors 
$\{ |0\rangle |0\rangle ,
|0\rangle |1\rangle \}$ and the state $|\Psi_{2out}\rangle$ lies in the 
space spanned by
$\{ (|0\rangle |0\rangle + |1\rangle |1\rangle ), (|0\rangle |1\rangle 
+|1\rangle |0\rangle ) \}$.  That
means that distinguishing $|\Psi_{1out}\rangle$ and $|\Psi_{2out}\rangle$ 
reduces to the problem of
distinguishing these two supspaces.  Making the correspondence with our 
previous example
\begin{eqnarray}
|0\rangle |0\rangle \rightarrow |0\rangle & \hspace{1cm} & |1\rangle 
|1\rangle \rightarrow |2\rangle
\nonumber \\
|0\rangle |1\rangle \rightarrow |1\rangle & \hspace{1cm} & |1\rangle 
|0\rangle \rightarrow |3\rangle ,
\end{eqnarray}
we see that the problem reduces to the one we have already solved.  
The subspace in which
$|\Psi_{1out}\rangle$ lies corresponds to $S_{1}$ and the one in which 
$|\Psi_{2out}\rangle$ lies
corresponds to $S_{2}$.  Therefore, the POVM we have already found will 
optimally distinguish (assuming that the input state is $|0\rangle |0\rangle$)
 through which black box the input state was sent.

\section{Conclusion}\label{section4}
We have presented a POVM that optimally and unambiguously 
discriminates between two
subspaces.  The construction of this POVM made use of the Jordan bases of 
the two subspaces. The results are, in fact, more general than what is 
stated in the title. They represent the complete solution to the problem 
of optimal unambiguous discrimination between mixed states of a special 
class, viz. between those states for which the spectral form coincides with 
the Jordan representation.
  
We presented two applications of the measurement procedure, 
discriminating between two-particle states if one has only one of the 
particles and deciding to which of two sets an unknown quantum operation
belongs.

This procedure can be used to distinguish arbitrary mixed states by 
discriminating between their
supports, but the results will not, in general, be optimal.  In order 
to optimally discriminate
between mixed states the structure of the states within their supports 
must be taken into account. Based, however, on the results of this paper 
we believe that this is an extremely difficult task for density matrices of 
Rank $>2$. The case of optimally discriminating between arbitrary Rank $2$ 
density matrices appears more tractable, however.  How it can be accomplished 
is a problem that still remains open. 

\begin{acknowledgments}
This research was partially supported by a grant from PSC-CUNY as well 
as by a CUNY collaborative grant. JB gratefully acknowledges many helpful 
discussions with Ulrike Herzog (Humboldt University, Berlin), on various 
aspects of state discrimination.
\end{acknowledgments}


\begin{thebibliography}{99}
\bibitem{springer}J.\ A.\ Bergou, U.\ Herzog, and M.\ Hillery, Lect. Notes Phys. {\bf 649}, 417-465 (Springer, Berlin, 2004). 
\bibitem{bennett} C.\ H.\ Bennett, \prl {\bf 68}, 3121 (1992).
\bibitem {bergou1}J.\ A.\ Bergou, U.\ Herzog, and M.\ Hillery, \prl {\bf 90}, 257901
(2003).
\bibitem{helstrom}C.\ W.\ Helstrom, {\it Quantum Detection and Estimation Theory} 
    (Academic Press, New York, 1976). 

\bibitem{ivanovic}I.\ D.\ Ivanovic, Phys.\ Lett.\ A {\bf 123}, 257 (1987).
\bibitem{dieks}D.\ Dieks, Phys.\ Lett.\ A {\bf 126}, 303 (1988).
\bibitem{peres}A.\ Peres, Phys.\ Lett.\ A {\bf 128}, 19 (1988).
\bibitem{jaeger}G.\ Jaeger and A.\ Shimony, Phys.\ Lett.\ A {\bf 197}, 83 (1995).
\bibitem{bergou2}J.\ A.\ Bergou, M.\ Hillery, and Y.\ Sun, J.\ Mod.\ Opt.\ {\bf 47}, 487 (2000).
\bibitem{terno}A.\ Peres and D.\ Terno, J.\ Phys.\ A {\bf 31}, 7105 (1995).
\bibitem{sun1}Y.\ Sun, J.\ A.\ Bergou, and M.\ Hillery, \pra {\bf 64}, 022311 (2001).
\bibitem{chefles}A.\ Chefles, Phys.\ Lett.\ A {\bf239}, 339 (1998).  See also A. Chefles, Contemporary Physics {\bf 41}, 401 (2000).
\bibitem{zhang}S.\ Zhang and M.\ Ying, \pra {\bf 65}, 062322 (2002).
\bibitem{sun2}Y.\ Sun, J.\ A.\ Bergou, and M.\ Hillery, \pra {\bf 66}, 032315 (2002).
\bibitem{rudolph}T.\ Rudolph, R.\ W.\ Spekkens, and P.\ S.\ Turner, \pra {\bf 68},
010301(R) (2003).
\bibitem{feng}Y.\ Feng, R.\ Duan, and M.\ Ying, \pra {\bf 70}, 012308 (2004).
\bibitem{raynal}P.\ Raynal, N. L\"{u}tkenhaus, and S.\ van Enk, \pra {\bf 68}, 022308 (2003).
\bibitem{HB} U.\ Herzog and J.\ A.\ Bergou, \pra {\bf 71}, 050301(R) (2005).
\bibitem{raynal2} P.\ Raynal and N.\ L\"{u}tkenhaus, quant-ph/0502165.
\bibitem{bergou3} J.\ A.\ Bergou and Mark Hillery, Phys.\ Rev.\ Lett.\ 
{\bf 94}. 160501 (2005).
\bibitem{jordan} P. X.\ Gallagher and R. J.\ Proulx, in {\it Contributions to
Algebra}, Bass, Cassidy, and Kovacic eds. (Academic Press, New York, 1977).
\bibitem{acin} A.\ Acin, Phys.\ Rev.\ Lett.\ {\bf 87},177901 (2001).
\bibitem{arianao} G.\ M.\ D'Ariano, P.\ L.\ Presti, and M.\ G.\ A.\ Paris
J.\ Opt.\ B {\bf 4} 273 (2002).
\bibitem{chefles2} A.\ Chefles and M.\ Sasaki, Phys.\ Rev.\ A {\bf 67}, 032112
(2003).
\end{thebibliography}
\end{document}